\documentclass[a4paper]{article}

\usepackage{graphicx}
\usepackage{color}
\usepackage{graphicx}
\usepackage{txfonts}
\def\apj{ApJ}
\def\apjl{ApJL}

\def\mnras{MNRAS}
\def\aj{AJ} 
\def\aap{A\&A} 
\newcommand{\Mpc}{$h^{-1}$\thinspace Mpc}

\begin{document}

{BIASING PHENOMENON\footnote{Paper presented at the Fourth
    Zeldovich meeting, an international conference in honor of
    Ya. B. Zeldovich held in Minsk, Belarus on September 7--11,
    2020. Published by the recommendation of the special editors:
    S. Ya. Kilin, R. Ruffini and G. V. Vereshchagin.}}\\

{\bf Jaan Einasto} \\

{\em Tartu Observatory, 61602 T\~oravere, Estonia}\\

\abstract {We study biasing as a physical phenomenon by analysing
  power spectra (PS) and correlation functions (CF) of simulated
  galaxy samples and dark matter (DM) samples. We apply an algorithm
  based on the local densities of particles, $\rho$, to form populations of
  simulated galaxies, using particles with $\rho \ge \rho_0$.  We
  calculate two-point CF of projected (2D) and spatial (3D) density fields of
  simulated galaxies  for various particle-density limits
  $\rho_0$. We compare 3D and 2D CFs; in 2D case we
  use samples of various thickness to find the dependence of 2D CFs on
  thickness of samples.  Dominant elements of the cosmic web are
  clusters and filaments, separated by voids filling most of the
  volume.  In individual 2D sheets positions of clusters and filaments
  do not coincide. As a result, in projection clusters and filaments
  fill in 2D voids. This leads to the decrease of amplitudes of CFs in
  projection.  For this reason amplitudes of 2D CFs are lower than
  amplitudes of 3D CFs, the difference is the larger, the thicker are
  2D samples.  Using PS and CFs of simulated galaxies and DM we
  estimate the bias factor for $L^\ast$ galaxies,
  $b^\ast =1.85 \pm 0.15$.  }


\onecolumn
\nopagebreak[4]
\section{Introduction}

\begin{figure*}[ht]
\centering 
\hspace{2mm}  
\includegraphics[width=0.44\textwidth]{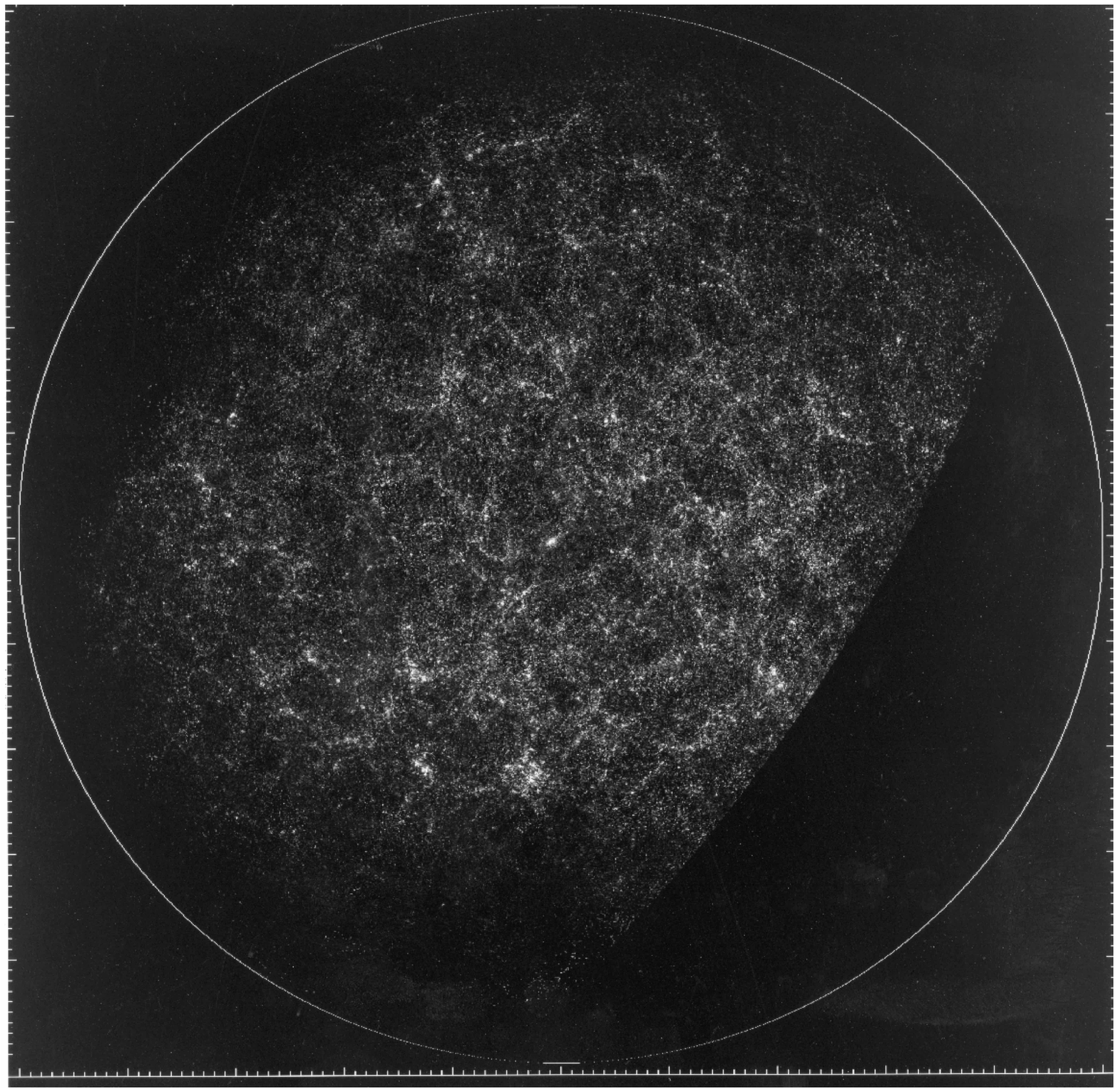}
 \hspace{2mm}  
\includegraphics[width=0.46\textwidth]{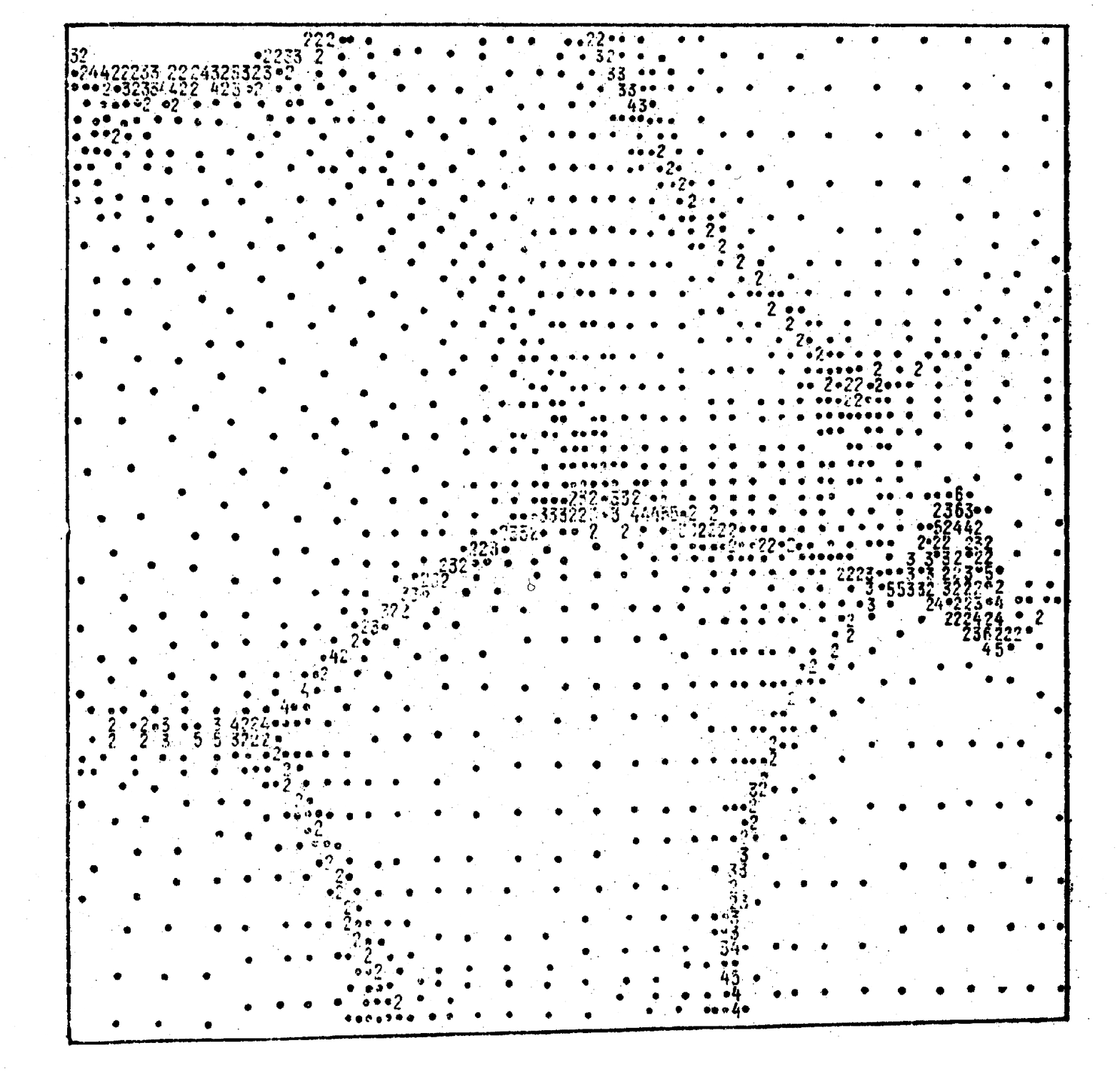}\\
\hspace{2mm}
\caption{Left: Distribution of galaxies according to Lick survey,
  reduced by Soneira and Peebles \cite{Soneira:1978fk}.  Right:
  distribution of particles in 2D numerical simulation by Doroshkevich
  and Shandarin \cite{Doroshkevich:1980}.}
\label{EinastoJfig1}
\end{figure*}

The biasing problem is as old as our understanding on the presence of
the 
cosmic web. Early data on the distribution of galaxies were obtained
from counts of galaxies on the sky. The most complete counts were made in
Lick Observatory, and reduced by Soneira and Peebles \cite{Soneira:1978fk}, see
Fig.~\ref{EinastoJfig1}.  This survey shows the presence of numerous
clusters of galaxies and a few superclusters, most of galaxies belong
to an essentially random  field population, which fills most the
whole sky in projection.

\begin{figure*}[ht]
\centering 
\hspace{2mm}  
\includegraphics[width=0.50\textwidth]{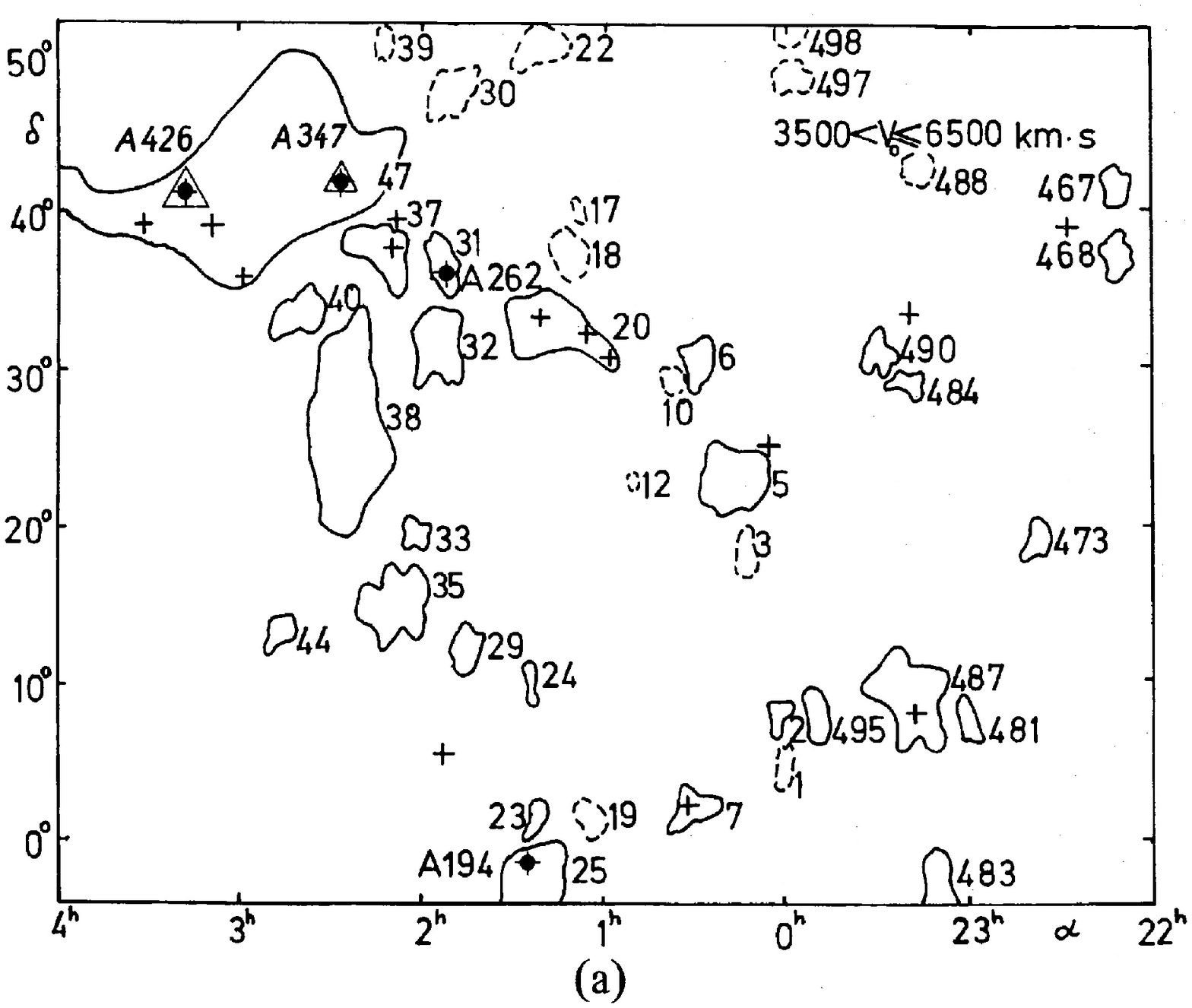}
\hspace{2mm}  
\includegraphics[width=0.42\textwidth]{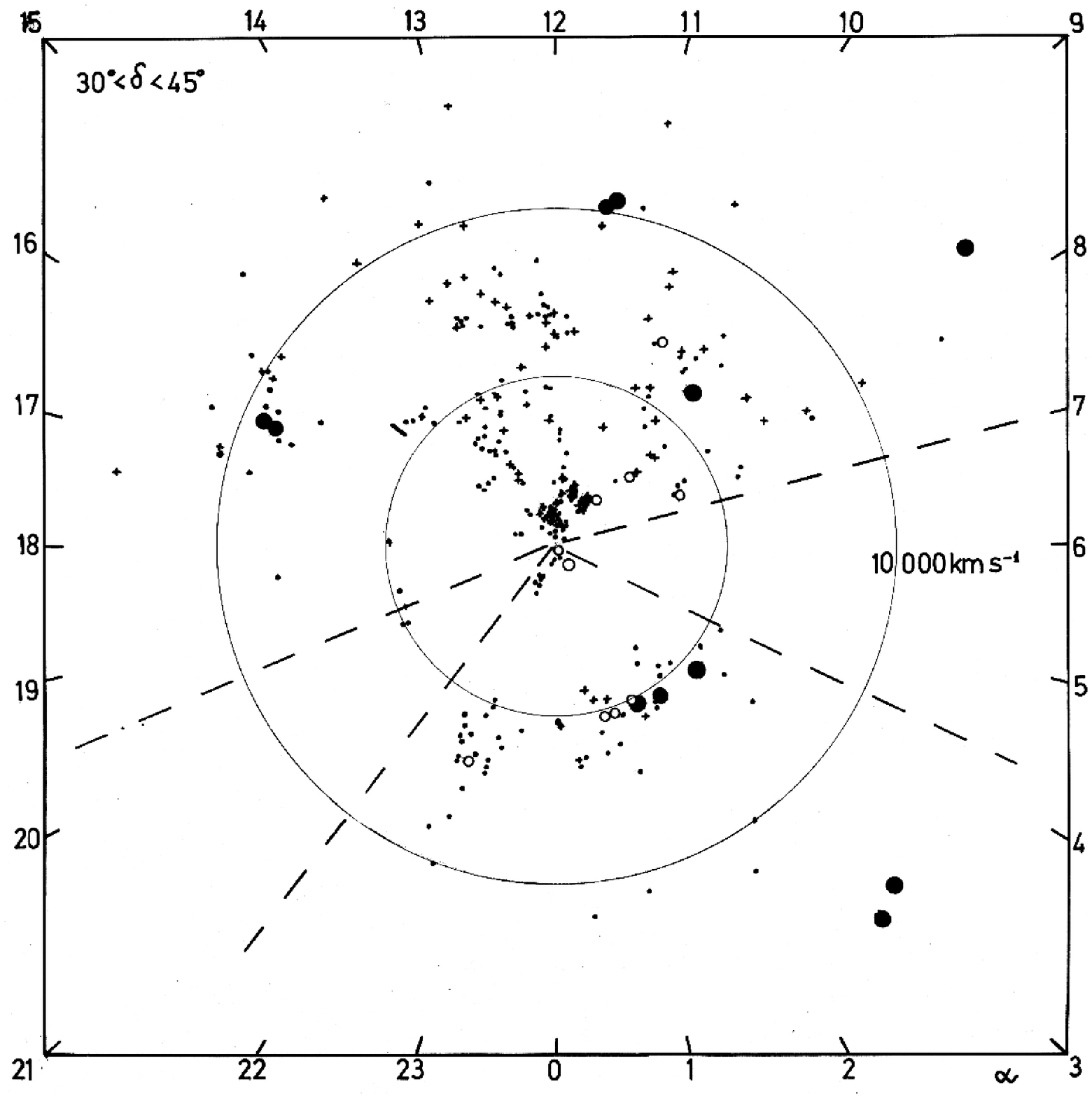}\\
\caption{Left: The distribution of Zwicky clusters of the distance
  class near in the Perseus area of the sky in redshift interval 3500
  to 6500 km/s \cite{Joeveer:1977}. Right: distribution of galaxies in
the Perseus-Pisces region \cite{Joeveer:1978pb}. }
\label{EinastoJfig2}
\end{figure*}

In mid 1970s the number of galaxies with known redshifts was large
enough to find the spatial distribution of galaxies.  New
three-dimensional (3D) data showed that in the distribution of
galaxies there exists a certain pattern, called presently the cosmic
web. Galaxies and clusters of galaxies are concentrated to filaments
and chains, the space between filaments is almost empty, as shown by
J\~oeveer and Einasto \cite{Joeveer:1977}, \cite{Joeveer:1978pb} in
the IAU Symposium on Large Scale Structure of the Universe, see
Fig.~\ref{EinastoJfig2}.  A theory of galaxy formation was suggested
by Zeldovich \cite{Zeldovich:1970}.  First two-dimensional (2D)
numerical simulations, based on Zeldovich model, were made by
Doroshkevich and Shandarin \cite{Doroshkevich:1976aa},
\cite{Doroshkevich:1980}, see Fig.~\ref{EinastoJfig1}.  Simulations
show the formation of the cosmic web with rich superclusters,
connected by filaments, and large voids, populated by a rarefied field
of test particles. This picture is similar to the observed
distribution of galaxies. However, there is an important difference.
In the simulation model there exist test particles in low density
regions, in the real universe voids are almost empty. On the other
hand, as noted by J\~oeveer and Einasto \cite{Joeveer:1977}, {\em it
  is very difficult to imagine a process of galaxy and supercluster
  formation which is effective enough to evacuate completely such
  large volumes as cell interiors are.}  There must be invisible dark
matter (DM) in voids.  This was the first hint to differences in
distributions of mass (DM) and galaxies.  Differences between
distributions of galaxies of different luminosity and clusters of
galaxies were studied by Kaiser \cite{Kaiser:1984}, who used the
term ``biasing'' for this phenomenon.

We consider biasing as a physical phenomenon, concerning the relation
between distributions of matter and galaxies on large scales. In
discussing the global biasing phenomenon we assume that gravity is the
dominating force,  determining the formation and evolution of the cosmic
web on large scales.  We consider the bias function $b$ as a function,
which quantitatively relates differences between distributions of
matter and galaxies. The numerical value of the bias function can be
found by the power spectrum (PS) and the correlation function (CF)
analyses.  We shall compare PS and CFs of simulated galaxy samples and
full samples of particles in models of the evolution of the cosmic
web.  The use of simulated samples of galaxies has the advantage that
all data are available for galaxies and matter.

\section{Biasing model}

Already first data on the distribution of galaxies and matter,
discussed above, suggested  that biasing is a threshold phenomenon --- in
low-density regions there are no galaxies. For this reason we cannot
use a simple biasing model, $b =\rho_g(X)/\rho_m(X)$, where
$\rho_g(X)$ and $\rho_m(X)$ are densities of galaxies and matter on
location $X$, respectively. Instead we use a biasing model, where
galaxies do not form in low-density regions at all, or are too faint
to be included into flux-limited galaxy surveys.  We label all
particles of $\Lambda$CDM simulation with local density value, $\rho$,
and apply a sharp particle-density limit, $\rho_0$. Simulated galaxy
samples consist of particles with density labels, $\rho \ge
\rho_0$. Full DM samples consist of all particles, i.e. particles with
density limit  $\rho_0=0$.

We shall describe the biasing phenomenon  quantitatively by the bias
function $b$, which can be calculated from the CF,
\begin{equation}
  b(r, \rho_0) = \sqrt{\xi_C(r,\rho_0)/\xi_m(r)}, 
\label{biasCF}  
\end{equation}
where $\xi_C(r,\rho_0)$ is the CF of clustered
matter --- galaxies, and $\xi_m(r)$ is the CF of all
matter, $r$ is the separation of particles/galaxies, and $\rho_0$ is
the  matter density limit, which selects 
 particles to be included. Another possibility is to use the PS,
$b(k,\rho_0)=\sqrt{P_C(k,\rho_0)/P_m(k)}$,
where $P_C(k,\rho_0)$ is the PS of true or simulated
galaxies, $P_m(k)$ is the PS of mass, and $k$ is wavenumber.

Both the CF and the PS methods have been used to study the clustering
of galaxies. Norberg \cite{Norberg:2001aa} investigated the clustering
of galaxies of the 2dF Galaxy Redshift Survey. Tegmark et
al. \cite{Tegmark:2004aa} and Zehavi et al. \cite{Zehavi:2011aa}
determined power spectra of galaxies of the SDSS survey, see
Fig.~\ref{EinastoJfig4}. To avoid redshift distortions authors
calculated as first step 2D CF of galaxies.  These studies allowed to
determine with good accuracy relative bias functions of galaxies of
various magnitude. Authors found that galaxies
approximately follow matter.  In our view this result is difficult to
accept, since already a visual comparison of distributions of real and
simulated galaxies with the distribution of matter shows large
differences. To understand the reason for this difference in global
bias parameter it is needed to study in more detail properties of 2D
CF.

The main difference between distributions of matter and galaxies is
the presence of DM in low-density regions, with no corresponding
population of galaxies. Thus a straightforward application of the equation
(\ref{biasCF}) is difficult, since there is no observable population
of DM.  We shall overcome this difficulty by using a model
distribution both for simulated galaxies and matter (DM). We assume
that modern models of $\Lambda$ cold dark matter ($\Lambda$CDM)
represent the real universe well.  Further we apply a threshold
clipping biasing model, described above.  Also we apply a new approach
to measure CF and PS --- the density field method by Szapudi et
al. \cite{Szapudi:2005aa}. In this approach the input information is
identical for CF and PS analysis -- high-resolution density field.
This approach allows to have a new look to CF, similar to the PS
analysis. Both start from density fields, use FFT, PS analysis finds
the dependence of bias function on wavelength scale $k$, CF on
relative spatial coordinates, $r$.  In both analyses we used a
$\Lambda$CDM simulation with cube size $L_0 =512$~\Mpc, and number of
particles $N_{\mathrm{part}}= 512^3$, assuming cosmological parameters
$\Omega_{\mathrm{m}} = 0.28$, $\Omega_{\Lambda} = 0.72$,
$\sigma_8 = 0.84$, and the dimensionless Hubble constant $h = 0.73$.

\section{Correlation  analysis of $\Lambda$CDM universe}

Here we analyse properties of 3D and 2D CFs of the $\Lambda$CDM model,
based on Einasto et al. \cite{Einasto:2020aa} and
\cite{Einasto:2020ab}. The analysis of PS of the $\Lambda$CDM model by
Einasto et al. \cite{Einasto:2019aa} yields similar results for the
biasing problem. In both analyses the same $\Lambda$CDM simulation in
a cube of size $L_0 =512$~\Mpc\ was used.  Simulated galaxy samples
with particle density limits $\rho_0$ were selected with
$\rho_0 = 1,~2, 5,~\dots , 100$, and with $\rho_0=0$, i.e. full DM
model sample.  Einasto et al. \cite{Einasto:2019aa} compared spatial
density fields of simulated galaxy samples and galaxy samples from
Sloan Digital Sky Survey (SDSS).  This comparison shows that density
fields of simulated galaxies with particle density limits
$\rho_0 \approx 5$ are similar to density fields of SDSS galaxies of
absolute magnitude $M_r \approx -18$.  $L^\ast$ galaxies of magnitude
$M_r =-20.5$ have spatial distribution properties, similar to spatial
distribution of simulated $\Lambda$CDM samples with density threshold
$\rho_0=10$.

\bigskip
\begin{figure*}[ht]
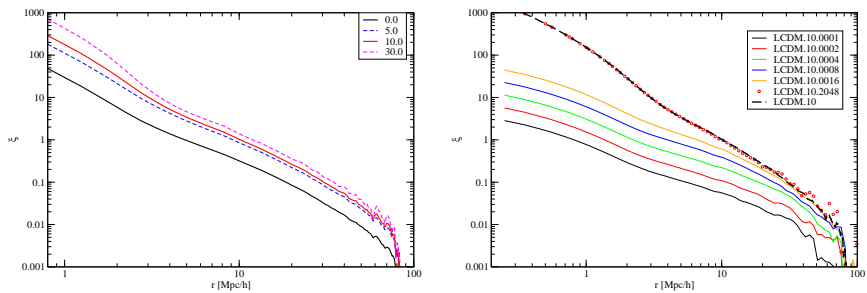

\centering 
\hspace{2mm}  
\includegraphics[width=0.45\textwidth]{EinastoJ_Fig5.eps}
\hspace{2mm}  
\includegraphics[width=0.45\textwidth]{EinastoJ_Fig6.eps}\\
\caption{CFs of simulated galaxies in
$\Lambda$CDM universe: left in 3D, right in 2D \cite{Einasto:2020aa}, \cite{Einasto:2020ab}. }
\label{EinastoJfig3}
\end{figure*}

\begin{figure*}[ht]
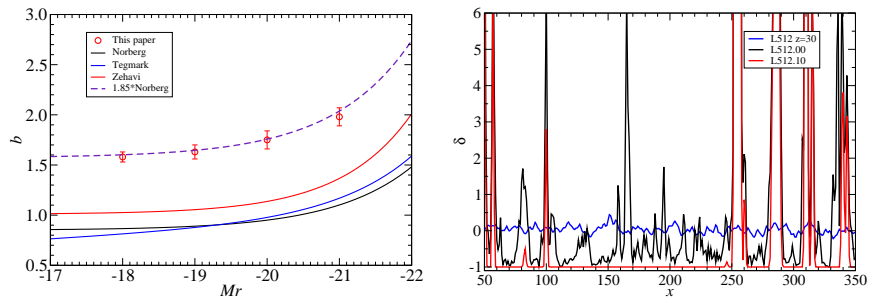

\centering 
\hspace{2mm}  
\includegraphics[width=0.45\textwidth]{EinastoJ_Fig7.eps}
\hspace{2mm}  
\includegraphics[width=0.45\textwidth]{EinastoJ_Fig8.eps}\\
\caption{Left: Bias function found by Einasto et
  al. \cite{Einasto:2019aa}  is marked by red circles with error
  bars. Black, blue, and red lines show bias functions  by Norberg et al. 
 \cite{Norberg:2001aa},  Tegmark et al. \cite{Tegmark:2004aa}, and
 Zehavi et al. \cite{Zehavi:2011aa}. Dashed 
  line shows the fit by Norberg et al. \cite{Norberg:2001aa}, applying bias
  normalising factor $b_\circ=~1.85$.
  Right: Cross section of density fields: blue line is for early
universe at redshift $z=30$, black line is for DM at the present
epoch, red line corresponds to matter in galaxies at the present
epoch \cite{Einasto:2019aa}.
}
\label{EinastoJfig4}
\end{figure*}

In Fig~\ref{EinastoJfig3} we show in the left panel 3D CFs of
simulated galaxies in our $\Lambda$CDM universe. Solid bold line
corresponds to the CF of all matter with particle density threshold
$\rho_0=0$, blue dashed line is for the sample $\rho_0=5$, red solid
line for the sample $\rho_0=10$, and violet dashed line for the sample
$\rho_0=30$.  These coloured lines correspond to real galaxies of
absolute red magnitudes $M_r = -18,~-20.5,~-22.0$.  We see that all 3D
CFs are approximately parallel, see Fig.~\ref{EinastoJfig4}.  As shown by Einasto
et al. \cite{Einasto:2019aa}, the percolation analysis suggests for
galaxies of absolute magnitude $M_r=-19.0$ the bias parameter
$b=1.68 \pm 0.09$, and for galaxies of absolute magnitude $M_r=-20.5$
the bias parameter $b=1.85 \pm 0.15$. This magnitude corresponds to
$L^\ast$ galaxies.

In the distant observer approximation we can calculate projected 2D CFs
from 2D density fields, using model data in rectangular spatial
coordinates. First we calculated 2D density fields by integrating 3D
field by ignoring $z$-coordinates in calculation of density fields in
the selected $z$ range.  We made the integration of cubical samples to
get $n$ sequentially located 2D sheets of thickness $T = L_0/n$, with
$n = 1,~2,~4, \dots 2048$.  $n=1$ corresponds to the whole sample in
$z$-direction of thickness, $T=L_0=512$~\Mpc, $n=2$ corresponds to
thickness $512/2 = 256$~\Mpc. For each $n$ we calculated 2D CFs for
all $n$ sheets, and then found the mean CF and its error for a given
$n$.  We show 2D CFs in the right panel of Fig.~\ref{EinastoJfig3} for
a sample of simulated galaxies with $\rho_0=10$, corresponding to
$L^\ast$ galaxies. Bold dashed black line shows 3D CF of this sample.
Solid coloured  lines correspond to 2D CFs of various $n$ and thickness.

Fig.~\ref{EinastoJfig3} shows that the integration of density fields
in the radial direction decreases the amplitude of 2D CFs.  2D CF of
the thickest slice with thickness $T=L_0/n=512$~\Mpc\ has at
separation $r=6$~\Mpc\ an amplitude, about 25 times lower than the 3D
CF of the same sample.  With increasing $n$ and decreasing thickness
$T$ the difference in amplitudes of 2D and 3D CFs decreases.  Only the
mean 2D CF for $n=2048$ and thickness $T=0.25$~\Mpc\ 2D CF is
identical to 3D CF of the same sample.

\bigskip

\section{Discussion}

Right panel of Figure~\ref{EinastoJfig4} shows a cross section of the density field.
We plot here the density contrast $\delta(x) = \rho(x)-1$.  The blue line
shows the density contrast at the early epoch, corresponding to a
redshift of $z=30$.  At this early epoch, the amplitude of density
fluctuations is small, and density fluctuations are approximately
equal everywhere.  The black line gives the density contrast for the
same cross section at the present epoch for the sample with all
particles.  This density field is dominated by fluctuations of
the density contrasts in the range $-1 < \delta(x) < 2$, with a mean
around $\delta \approx -0.5$.  The red line shows the same cross
section for the simulated galaxy sample for $L^\ast$. Here
high-density peaks of the density field are the same as in the full
model, but weak DM knots of medium density, seen in the full DM sample,
are gone.  In contrast to the full sample, over most of the density
field the density of the simulated galaxy sample has zero density and
density contrast $\delta =-1$.

CFs of our models were calculated using density fields applying the
method by Szapudi et al. \cite{Szapudi:2005aa}.  As input for both
correlation and PS analysis is the density contrast field,
$\delta$, used for PS analysis as
follows:  $P(k)=\langle |\delta_k|^2\rangle$. In DM model there are
no cells with zero density, low-density cells have a mean density
contrast $|\delta| \approx 0.5$. In $L^\ast$ model sample most of these cells
are empty with $|\delta|= 1$. From the above equation it follows that this
increases the  amplitude of PS.  The increase of the amplitude is the
greater the larger is the fraction of zero density cells. The fraction
of zero density cells is 95\% of all cells both in the SDSS sample
and the $\Lambda$CDM model sample for $L^\ast$.  The PS and CF are
proportional to the sum over all density contrasts.  The greater the
fraction of cells with zero density and density contrast
$|\delta| = 1$, the greater the sum.  For this reason, the PS and CF of
all simulated galaxy samples have  higher amplitudes than the full DM
model; the larger the fraction of zero-density regions, the higher the
amplitude.

\begin{figure*}[ht]
\centering 
\includegraphics[width=0.75\textwidth]{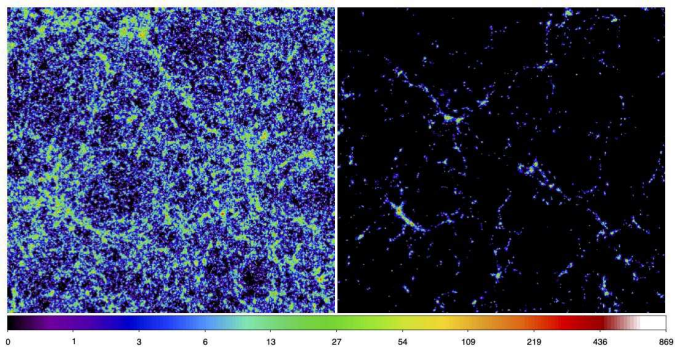}\\
\caption{2D density fields of various thickness. In left  panel a view of thickness
512~\Mpc, in right panel of thickness 32~\Mpc\ \cite{Einasto:2020ab}.
}
\label{EinastoJfig5}
\end{figure*}

We show in Fig.~\ref{EinastoJfig5} 2D density fields of the model
$\Lambda$CDM using particle density limit $\rho_0=10$, and having
various thickness. In left panel we present 2D density field of
thickness $T=512$~\Mpc, i.e. the whole cube of our $\Lambda$CDM model.  In
the right panel the thickness is $T=32$~\Mpc. Morphological properties
of 2D thin slices are close to properties of the 3D density field of
the same model: the fraction of high- and low-density regions, mutual
distances between high-density regions. Essential elements of the
cosmic web are high-density regions (clusters and filaments) and
voids, which together form the pattern of the web.  The detailed
structure of the web is seen in right panel of Figure.  High-density
regions are surrounded by zero density voids, occupying most of the
volume of the density field.  Clusters and filaments in sheets at
different $z$ locations are located in various $x,y$-positions.  In
stacked thick 2D sheets clusters and filaments at various $z$ are
projected to the 2D $x,y$ plane at different positions, and in this
way fill in voids in the 2D density field.  This is clearly seen in
the left panel of Fig.~\ref{EinastoJfig5}. In earlier studies this
effect was ignored. The lowering of the amplitude of 2D CFs can  be the
main reason why amplitudes of bias functions in earlier studies were
so low.

\section{Conclusions}

\begin{enumerate}
\item The presence in galaxy density field of large number of cells
  with zero density increases the amplitude of the galaxy PS and CF
  over the amplitude of matter PS and CF. Thus the amplitude of the
  galaxy PS and CF is always higher than the amplitude of PS and CF of
  matter, i.e. bias parameter is $b>1$. The bias parameter of $L^\ast$
  galaxies, as found from $\Lambda$CDM power spectra is $b=1.85 \pm 0.15$.

\item Dominant elements of the cosmic web are clusters and filaments,
  separated by voids filling most of the volume. In 2D sheets clusters
  and filaments are located at different positions fill in 2D voids,
  which leads to the decrease of amplitudes of CFs.  For this reason
  amplitudes of 2D CFs are lower than amplitudes of 3D CFs, the
  difference is the larger, the thicker are 2D samples.

\item Biasing problem is a surprisingly complex phenomenon, where
  theories of galaxy formation meet with studies of the structure of
  the cosmic web. The use of the density field in CF and PS analyses
  allows to see these functions from a different point of view.
\end{enumerate}
\bigskip

  I thank my coauthors in the study of the cosmic web Maret Einasto,
  Gert H\"utsi, Juhan Liivam\"agi, Ivan Suhhonenko, Elmo Tempel for
  fruitful and enjoyable collaboration, and  Enn Saar for
  discussion.


\section*{Funding}

 This work was supported by institutional research funding IUT40-2 of
  the Estonian Ministry of Education and Research, by the Estonian
  Research Council grant PRG803, and by Mobilitas Plus grant MOBTT5. We
  acknowledge the support by the Centre of Excellence``Dark side of
  the Universe'' (TK133) financed by the European Union through the
  European Regional Development Fund.

\end{document}